# Coexisting Spacetimes in the Solar Neighborhood

Alasdair Macleod
University of the Highlands and Islands
Lews Castle College
Stornoway
Isle of Lewis
HS2 0XR
UK

Alasdair.Macleod@lews.uhi.ac.uk

*Abstract* We consider the proposition that multiple universes exist by reviewing the various manifestations. In recent years, this idea has been elevated from science fiction and introduced in separate guises as an explanation for coincidence problems in cosmology, the prediction of dark energy, gravitational anomalies, a consequence of string theory, an extension of inflation, and conceptual issues in quantum mechanics. However, there appears to be no single consistent formalism that addresses all the issues – it is not even clear if the multiple universes interact or are accessible. Because of the absence of clear evidence, it is easy to dismiss claims for multiple Universes, but it is not that simple: Space-time geometries in a variety of forms are an established aspect of physics, and history has shown that things which are not expressly forbidden often appear at a later date. With this in mind, a new example of possible multiple geometries is introduced to explain a difficult problem in cosmology: Why are distant galaxies subject to the Hubble expansion but rulers within our galaxy (and presumably, by the Cosmological Principle, all other galaxies) do not appear to expand? There has actually been much debate in recent years about whether local systems really are subject to the cosmological expansion; we adopt the established view that local systems are not expanding and investigate the transition condition where a mass may be part of the static galactic system or participate in the expansion. We show that a remarkably simple model based on coexisting spacetimes clarifies the situation and makes quantative, testable cosmological predictions. The model is also testable through Earth-based experiments.

## 1. INTRODUCTION

The theoretical development of General Relativity (GR) from its inception up to the end of the $20^{th}$ century converged on a beautifully simple mathematical model for the Universe as a closed system curved by the effect of its own mass. However, recent observations have shown that the real Universe deviates significantly from the model. The mainstream response has been to complicate the model by introducing additional fields and entities to bring data and model into agreement. This is not the only approach – it is possible to retain the concept of a simple 'tidy' Universe but to postulate that a multitude of such Universes exist. The deviant effects then arise in the way these Universes interact. This is a very elegant idea but not original – the concept is possibly borrowed from Quantum Theory where the many-worlds interpretation had decades earlier helped clarify difficult conceptual problems at the heart of the subject (for some). Although impossible to visualise, but surely no worse than curved spacetime and string theory, the idea of multiple Universes, or *Multiverses*, are strongly supported by the reasonable enquiry, "If one Universe can exist, why should the creation process not repeat itself over and over again?" In fact, the only way to exclude the notion of multiple interacting Universes is to demonstrate that 'our' Universe is a closed and logically complete system in its interactions.

Multiverses have also been proposed as a scientific response to the incredible fine-tuning required to give rise to a Universe capable of supporting life (the anthropic principle). Tegmark provides a good review of the subject [1].

The main criticism of the Multiverse concept in any of its forms is that there is little or no evidence for these other Universes [2], but if a reasonable model based on Multiverses makes correct predictions, this may be interpreted as suggestive of their existence [3]. The great difficulty in constructing a compelling theory is deciding how the Universes should interact – there is simply no guidance available.

With so few constraints other than elegance, the floodgates are opened with the result that we are bombarded with alter-Universes, for example, mirror universes, shadow universes and parallel universes[i], none more unreasonable than the other. One has to eventually query which of the emerging ideas represent genuinely science, but how can one judge? Some decades ago, such speculations would never have been entertained, but the deep mystery of dark matter and dark energy is driving science in more and more speculative directions. We will look at a representative sample of ideas on multiple Universes (taken only from those that have been gained some scientific recognition) and judge if they contribute or complicate- in other words, is the particular Multiverse model more complex than the alternative, a single misbehaving Universe.

A puzzling feature of the cosmological expansion is why it does not appear to operate on the scale of galaxies or smaller. We consider this in the context of Multiverses and demonstrate that a dual spacetime model clears some conceptual issues. Unusually for this field, these spacetimes can interact through the shared matter content and give rise to observable effects. The predicted effects that emerge naturally would be considered anomalous in a single spacetime model.

## 2. QUANTUM MULTIVERSES

The first significant theory to incorporate multiple Universes was Everett and DeWitt's many-worlds interpretation (MWI) of quantum mechanics [4]. When quantum systems interact, they proposed that the outcome is a set of altered Universes, all independent and non-interacting, differing only in the quantum state following the measurement event. All possible measurement outcomes are represented once by the set of Universes that result. The observer exists in each of these Universes and is duplicated through the branching process[ii]. The theory arose through dissatisfaction with the conventional Copenhagen interpretation of quantum mechanics, which considers the observed state following an observation (the collapsed function eigenstate) to be intrinsically random. The MWI removes the *ad hoc* nature of wave function collapse, as there is now no preferred state; all are represented in the daughter Universes. Determinism is fully restored.

It is difficult to do justice to what is a very imaginative idea in a brief review, but the MWI appears mathematically and conceptually sound. The interpretation has been criticised, but mainly on aesthetic grounds or on how the human experience is maintained (apparently as a single thread) through the branching process. However, these are metaphysical not physical issues. Of greater interest here is how the MWI deals with another problem in Quantum Theory, the apparent need for acausal or non-local activity at a quantum level in contrast to the macroscopic level where instantaneous action-at-a-distance is prohibited by the theories of relativity.

There is currently no consensus: One of the claims made for the MWI is that it is a local theory [5, 6], but it is also claimed by some to be a non-local theory [7]. If non-locality is identified with any requirement for information transfer at a speed faster than that of light, then experiments that explore the collapse event certainly show that Quantum Theory is fundamentally non-local: Entangled photons collude in a way that appears to violate causality [8]. The MWI does not alter this conclusion. In spite of claims to the contrary, it is easy to show that the MWI is actually a non-local interpretation of quantum mechanics. We may demonstrate this through an operational approach, by asking about the mechanism through which the MWI model operates. Actions in the MWI take place in phase space. Effectively, the Universe, represented by quantum state vectors, replicates on each photon transfer. When we translate this process to a space viewpoint, we can legitimately ask how the copies are made. At the moment of measurement, knowledge over the vast extent of spacetime is required instantly to create the copies[iii]. This is a pretty good example of a non-local process.

The MWI is therefore not an explanation for quantum non-locality. The issue of non-locality can hardly be ignored – if non-locality just happens to be a feature of our world that we have to accept, why does it not operate at a macroscopic level? For example, many attempts have been made to develop Mach's principle[iv] through action-at-a-distance influences but in all cases other effects are predicted that have never been observed [9]. In any case, the issue of non-locality sits very uncomfortably with relativity.

There are other operational problems as well; we might ask about the instances where two or more events occur simultaneously[v] - how does the branching then take place? Do measurement processes conveniently schedule themselves at separate intervals and thus somehow determine the progression of time? This is a pretty far-fetched notion; how could such control be exerted? Of course with the full MWI formalism, there are workarounds based on only parts of the Universe being replicated (worlds are considered to be closed sub-systems), but how then is the boundary of the region to be determined [10]?

In terms of taxonomy, the MWI is only one of a number of philosophically related ideas, all embracing non-locality. Although not strictly examples of Multiverses, interesting variants have been proposed, also in the context of quantum mechanics, with the core idea of the Universe at a future time being able to directly influence on the Universe at the present time. The end result is the same as with the MWI, non-locality. The two main models of this type are the Feynman-Wheeler absorber theory [11] and Cramer's transactional interpretation [12]. They generally do not alter the predictions of quantum mechanics but are purely interpretations of the energy exchange process with the objective of clearing the non-locality issue. The procedure in each case is described in terms of a field theory for incorporation into our existing understanding of the interaction process. The procedure works only if electromagnetic waves are permitted to travel backwards in time. But at a fundamental level, this is merely another description of non-locality[vi].

These and similar ideas [13, 14, 15] are speculative and ultimately converge on a picture of the Universe as a unified whole, extended in both space and time. The idea that interactions have proximate causes is then illusionary. The system is complete and consistent – photons do not wander about until they are absorbed, they are simply representations of energy transfer. In term of our current ontology, photons are emitted *because* the absorber will be there, thus the notion of propagation has no meaning; there is no choice of destinations – in fact there is no choice at all. Fields and a substantive space become unnecessary. The Universe is completely deterministic[vii]. Theories of this type are metaphysically attractive but are of limited interest to the physicist because of the massive questions that arise. For example, the transition probability from perturbation theory (applied to Dirac's equation) is derived correctly from only the local fields without reference to future absorbers. Also, the twin-slits experiment demonstrates that the space path is a factor in determining the probability of a photon being absorbed, hence propagation would seem important and cannot just be ignored.

The MWI and all the variants are certainly interesting and worthy of study, but ultimately do not solve all the interpretation problems associated with quantum mechanics.

## 3. INFLATION AND ANTHROPIC MULTIVERSES

The evidence at this stage is that the Universe is flat: The density of its content is equal to, or very close to, the critical value that defines the transition between the closed and open Universes emerging as solutions to Friedmann's Equations. This is fine as a working definition, but for conceptual purposes it is useful to relate the idea to geometry. In the classic expanding balloon model of the Universe [16] the geometrical curvature is non-zero and can theoretically be measured by drawing out a huge triangle with light beams and noting the sum of the interior angles. We would expect the total to deviate from 180 degrees. Although it is not technically feasible to conduct that particular experiment, the results can be inferred by other techniques. Surprisingly, the total is found to be exactly 180 degrees – the Universe is flat. The appropriate model is actually that of a flat rubber sheet being pulled from each corner. The challenge is to reconcile the expanding balloon with the flat sheet. That's where inflation comes in. Spacetime *appears* flat because we only perceive a small part of what is a huge balloon. This is only possible if the Universe expanded at some stage in its existence at a rate that was much greater that the speed of light. Thus the viewable Universe is much smaller that the actual Universe, the entire surface of the balloon, so to speak[viii].

The inflationary scenario was developed by Guth [18] and implies that there are causally disconnect regions in the Universe. Though not strictly a Multiverses model in its basic form, Linde extended the basic concept and postulated that inflationary expansion would continually take place by some sort of budding process to create a fractal Multiverse comprised of bubble Universes each possibly subject to different laws [19]. The process of growing new Universes might even take place within our own observable Universe, maybe, it has been suggested, inside black holes [20].

What do we make of all this? Implicit to these ideas is still the lack of interaction. It is unclear how one would observe these other Universes from ours. Nevertheless, there has been a surge of interest in these ideas because of the so-called fine-tuning problem. If, as is believed, the expansion of the Universe is a dynamic process where expansion is opposed by gravity and possibly other forces, the fine-tuning required for a flat Universe is staggering. Effectively a cosmological constant with a value differing from zero by an unbelievably tiny amount is required to model the Universe. This level of fine-tuning appears highly implausible since it makes our Universe special for no particular reason. In a sense this is an anthropic argument [21], and the idea that there are many Universes, each with different laws and values for force constants, restores some scientific credence. Our Universe is unusual, but with all the possible Universes, it is considered inevitable that one particular combination of circumstances will give rise to intelligent life.

The idea of a multitude of Universes is the only scientific answer on offer for the remarkably specific values of fundamental constants needed to create the stable Universe we inhabit. We argue here that concept is actually unscientific and cannot be justified. The problem is that the whole edifice is based on an assumption that is quite at odds with scientific progress. As scientific knowledge develops, what appear to be disparate effects tend to be unified under new physical models. In a sense, successful physical theories link ideas and reduce the degrees of freedom. There is no evidence that Universes (this Universe) are created with a set of unrelated initial parameters such as the mass of the electron, the cosmological constant and so on, that can take on any value (and rules that can take any form). The objective of scientific enquiry is to presume a relation exists and find the connection between them, in other words, find out how the characteristic numbers or ratios are associated. The idea of Multiverses as a solution to the anthropic problem smacks of numerology [23] and can only survive whilst we are ignorant of the relationship between fundamental parameters[ix]. One might imagine that a Theory of Everything, when it emerges, will show that the Anthropic Multiverse theory and its variants to be incorrect.

The difficulty of acquiring data about the other possible Universes hasn't stopped researchers exploring the possibilities. Ellis *et al* [22] rather remarkably describe the ensemble of all the actual Universes (which by their tighter definition are non-interacting) and apply statistical reasoning. However, merely expressing a concept in mathematical form does not mean mathematical methods can then be usefully applied to it.

There has been much criticism of the Multiverse explanations of the Anthropic principle, some of it fierce. We can conclude with a comment by Mosterin [23]:

> Alleged anthropic explanations do not explain anything and are not needed in cosmology. And if someone still intends to revive the corpse of anthropocentrism, he will need stronger medicine than just the anthropic principle itself.

### 4. STRING THEORY AND OTHER MULTIVERSES

String theory proposes that our perceived Universe is the projection onto three (or four) dimensions of a multidimensional world inhabited by string and membrane elements [25] Though not technically a Multiverse, presumably other projections and Universes are possible. For string theory, the inaccessible dimensions are troublesome and are dealt with by winding them up in a tiny way such that they become hidden. The current thinking in this field is that Universes can also arise from the collision of branes and that many types of Universes can result.

Whilst string theory is mathematically very interesting, it is very much an acquired taste[x]. The structure has so many degrees of freedom that it could predict anything, but ultimately predicts nothing. That is not to say it will not work – we are reminded here of the elaborate mechanical model used by Maxwell to successfully derive his field equations [26]. String theory has been criticised [27], but really the verdict is still open. However, the theory is certainly not so well-established that one can invoke brane-collisions as the origin of this Universe and others[xi].

There are many other examples of Multiverses, but these are even less successful that the ones reviewed in this paper. Although superficially intriguing, Multiverse models have tended to add little to our understanding of the world. For this reason, one must be very cautious about introducing Multiverse ideas. The minimum requirement is a clarification of our worldview in some way and some testable predictions.

### 5. LOCAL HUBBLE EXPANSION

Although the global expansion of the Universe is a well-established phenomenon, it is believed the space within galaxies does not expand [16, 28]. On the face of it, this distinction is odd, particularly as there is no conclusive empirical support for it. Neither is there any compelling theoretical justification. How did a seemingly arbitrary judgement become the accepted view? This is actually one instance where commonsense, not normally the scientist's most reliable guide, can be applied – if the expansion were ubiquitous and universal, local rulers would scale proportionally and the expansion would be undetectable. However, we have to be very careful with this type of argument and apply some rigour. The conclusion holds only if forces somehow cooperate - the expansion is only masked if the force constants change in a very specific way over time when measured from a non-expanding reference frame. We will see later that forces do seem to alter in the required way. Therefore, the fact that we can actually detect the cosmological expansion would indicate the Earth is a stable observatory that does not participate in the expansion.

The distinction between systems subject to the expansion and those that are not raises very difficult questions that have yet to be fully addressed. Much recent research has explored the alternative possibility that the expansion really does operate at a local level [29, 30, 31, 32]. These models get round the line of reasoning applied in previous paragraph by proposing that forces are unaffected by the expansion. One argument goes roughly as follows: Atomic particles (or even planetary positions) in a bound system are maintained through a balance of opposing forces. The expansion effect is locally very small but is certainly present and must be included as an effective outward force. As a result, the system equilibrates at a slightly extended position. Detailed calculation reveals that locally the change is miniscule and completely undetectable.

Although seemingly reasonable, the approach is ontologically unsatisfactory. It is not permissible to claim that forces are independent of the spacetime manifold over which they operate. To imply that forces have their own separate reference against which the spacetime expansion can be detected is unjustified and quite at odds with GR, a proven theory that after all identifies force with the topology of spacetime. We may conclude that in regions where the expansion is active, forces *will* collude with the expansion and modify with time in precisely the required way *because* forces are part of the expansion, thus, for once, the commonsense viewpoint is correct: Space does not locally expand otherwise we would not be able to detect the overall expansion. Misner, Thorne and Wheeler in their authoritative work simply state that stars within galaxies are not subject to the cosmological expansion [16]. One can only presume these issues seemed to them sufficiently intuitive that a mathematical justification for the conjecture was considered unnecessary.

Having defended the commonly held view of a galaxy immune from the expansion as the only sustainable ontology, we can discover some interesting consequences by exploring the transition state between expanding and 'static' space. It will be shown that the presence of two coexisting spacetimes here in our local solar environs is a logical requirement.

## 6. COEXISTING SPACETIMES

The accepted world model is that of General Relativity. Space is considered substantial - space would exist even in the absence of matter; furthermore, space is believed to be continuous. We will examine how space defined in this way can be made consistent with a Universe where some mass is subject to the cosmological expansion, other mass not.

Fig. 1 shows snapshots of two receding galaxies at time $T$ (Fig 1a) and a short time afterwards, $T + \Delta T$ (Fig 1b). We identify $T$ rather loosely with the time elapsed from the beginning of the Universe in the frame of the observer.

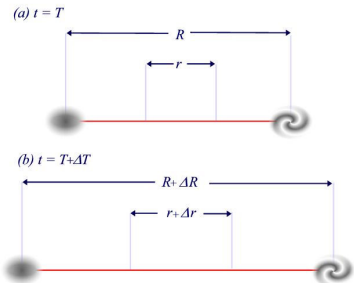

**Figure 1. Receding galaxies. The space marked $r$ increases to $r + \Delta r$ after $\Delta T$**

The portion of space of extent $r$ observed at time $T$ grows to a size $r+\Delta r$ after a time $T+\Delta T$. The change in length is consistent with an apparent velocity, $v$, described by the Hubble relation $v = Hr$, where $H$ is the Hubble constant. $H$ is determined from General Relativity. We assume the galaxies are so far apart that gravitational effects are negligible.

In Fig. 2 we look at the same situation but now a third galaxy is placed in the marked region (previously empty).

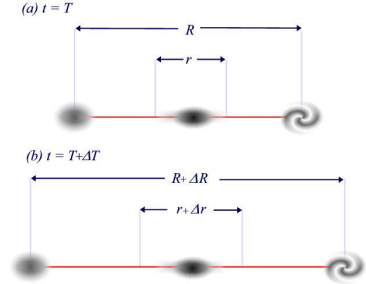

**Figure 2. Receding galaxies as before but with another galaxy occupying the marked position**

Again neglecting the gravitational effect, which can be made negligibly small, the outer galaxies move apart as before as must all the space between (the continuity condition). The same expanding spacetime that pushed the galaxies apart in Fig. 1 must also be present in Fig. 2 to maintain the dynamics of the expanding Universe (a process which does not exhibit local mass-energy dependence). The expanding spacetime connects through the newly introduced galaxy without affecting it. The space within this galaxy does not expand, thus we can identify a substantive local spacetime in which the stars that make up the galaxy are embedded and referenced. These two spacetimes are required for consistency and cannot possibly be the same entity. It therefore follows that in our galactic neighborhood (or in the neighborhood of any galaxy) two distinct spacetimes coexist. This conclusion follows inevitably from the postulates that space is substantial and the Hubble expansion does not act locally.

The single Universe with which we are familiar has been split into two distinct spacetimes. The *global spacetime* is subject to the cosmological expansion but the *local spacetime* is not. The extent and scope of each spacetime is undefined. The way we choose to understand these entities is very much dependent on what we consider space (and time) to be[xii]. Though tempting, it is incorrect to abstract these into mere frames against which matter is referenced because both spacetimes are present locally and could potentially be detected at the same time by an appropriate observational procedure. The spacetimes are considered capable of interaction through the shared matter content, with the nature and consequences of the interaction giving rise to observable effects.

Two spacetimes rather than one may seem like an unwelcome complication, but we would expect at the very least a clarification of the anomalies that currently arise, possibly because the Universe is incorrectly represented as a single spacetime. The decoupling process immediately confers some conceptual advantages. GR is currently applied in a different way to local systems than to the Universe as a whole. This approach is fully justified by the dual spacetime model. The local spacetime is essentially flat and can be treated special relativistically for the most part. The global spacetime must be treated general relativistically. The problem that some parts of the Universe expand and others do not is now completely resolved.

The role of time will not be discussed in any depth in this introductory paper. However, two (possibly coincident) time scales are implied. This is rather reminiscent of Milne's model of an infinite Universe with a kinematic time and a separate Newtonian time scale. Dirac also suggested that two time scales exist, an atomic and a global time [33]. Both these models predicted a variation in the value of the gravitational 'constant' and failed when no such variation was found by accurate ranging measurements within the solar system.

We can make some very general deductions about the relationship between the two frames from observation alone. Referring again to Fig. 2, the spacetimes appear linked at one point, presumably the centre of mass of the local frame. Each mass element is attached to one spacetime manifold only, though it can be observed from the other. The idea that a particle 'belongs' to a particular spacetime is important because we can investigate if and how the 'ownership' can be changed. This may then shed light on the detailed relationship between the spacetimes.

How does a particle or mass bind to one spacetime rather than the other and how is a transition between spacetimes effected? It is reasonable to identify the transition events with the movement between bound and free states. Bound masses or particles are locked into the local spacetime and free mass or particles are tied to the global spacetime. Thus if a mass gains or loses energy, for example through a collision, it is possible to move between spacetimes, as determined by the final energy. As a particle moves between spacetimes, we have to consider energy conservation. These are of course separate spacetimes and the concept of energy conservation between frames (as opposed to within frames) is at best poorly defined. There is no requirement that the spacetimes even respect the same energy scale, although we would expect there to be a strong connection between them.

Fig. 3 gives a general idea of how a mass might move between spacetimes. The diagram shows a mass (or particle) that is initially gravitationally bound receiving an energy boost and becoming free. Once free, the mass is 'attached' to the global spacetime manifold. The energy reference levels are arbitrarily defined and show that the apparent energy of the particle, once free from the bound reference frame is not necessarily $E_f$ when viewed by the gravitationally bound observer. It could be less than or greater than the expected value (it is less than in this example).

Once free the mass is subject to the cosmological expansion and will appear to recede from any local observer at a velocity that is the Hubble constant times the distance between the centre of mass and its new position[xiii].

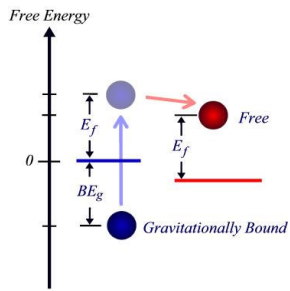

**Figure 3. Bound mass in blue (binding energy is $BE_g$) escaping and moving to the cosmological spacetime (red)**

There are known problems associated with energy conservation in GR [34], but these problems are ameliorated with the new formalism because the requirements for energy conservation can be considerably relaxed. We can demand energy conservation *within* spacetimes but not *between* spacetimes. Nevertheless we would expect there to exist a predictable and derivable relationship between the energy reference levels. Since it must be impossible to gain or lose energy by a closed loop process we can immediately deduce that both spacetimes have a shared energy reference at the linkage point.

To proceed further, we can generalise about how an unbound mass will appear in the local frame. The key again is consistency. The projection will be subject to the laws and conventions appropriate to the local frame although the host frame actually controls the dynamics. The mass in its own spacetime is subject an expansion of space governed by Friedmann's equations. In the global frame, the mass is not moving; instead space is expanding. The local flat spacetime knows nothing about expanding space and such considerations are irrelevant. The global effect is therefore projected onto the local frame as a proper velocity $v$. Photons received in the local frame will be Doppler-shifted to an extent that is consistent with the apparent velocity. The apparent velocity will remain constant since the local frame conserves energy. If it does not, then this indicates a (virtual) force in the local frame that can only arise from the interaction of the global and local frames.

The free particle or mass observed in the stationary frame can be treated kinematically and we will investigate if an energy relationship between spacetimes can be defined that will make the projection consistent and force-free.

Fig. 4 includes the total energy and shows a free mass that is apparently receding at a velocity $v_H$, the Hubble velocity. The diagram shows the reference at absolute time $T$ and a later time $T + \Delta T$.

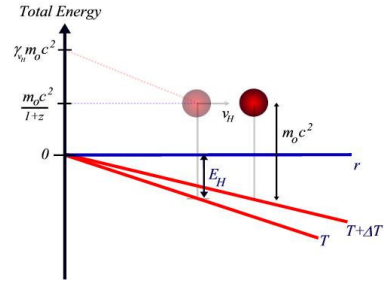

**Figure 4. A representation of a marginally free mass moving with a recession velocity $v_H$. The diagram shows the total energy at different times**

Assume the special case of a mass with zero proper velocity in the global frame. Because the mass is bound to the global frame rather than the local frame, the apparent rest energy is reduced by $E_H$. The observer will see the adjusted rest mass boosted by the Lorenz factor associated with the recession velocity. The measured energy is therefore dependent on $E_H$ but must be consistent with the redshift effect. We may therefore derive an energy difference function between frames that will guarantee consistency. This is developed below with reference to Fig. 4.

In time $\Delta T$, the mass has moved from $r$ to $r+v\Delta T$. For a mass subject to the cosmological expansion to appear to conserve energy and momentum, the following must hold:

$$v(r,T) = v(r + v\Delta T, T + \Delta T) \qquad (1)$$

The simplest particular solution is

$$v(r,T) = \frac{r}{T} (= Hr) \qquad (2)$$

Furthermore, for consistency with the redshifted energy, we can state that

$$\gamma_{v_H}(m_o c^2 - E_H) = \frac{m_o c^2}{1+z} \qquad (3)$$

Solving and using (2),

$$E_H = m_o v_H c = \frac{m_o r c}{T} \qquad (4)$$

Equation 4 defines a difference in the energy reference levels between the global and local spacetimes that ensures an unbound mass behaves consistently when observed from the local frame (as least to first order). Is this energy real[xiv]? This question is not easily answered. However, if the previous manipulation is more than a mere contrivance to reconcile general relativistic and special relativistic effects there should be measurable predictions associated with it. These predictions can be experimentally validated. However we have to derive very specific and measurable predictions in order to plan the appropriate experiments.

Fig. 5 shows a mass $m$ in a gravitational field bound in a stable circular orbit around a central mass $M$. There exists a condition where the mass can spontaneously become free and associate instead with the global spacetime. This will give rise to apparently anomalous behaviour when interpreted through our current worldview. For the centrally bound mass the binding energy is $-GMm/2r$ ($G$ is the gravitational constant). The diagram shows how the spontaneous transition can occur.

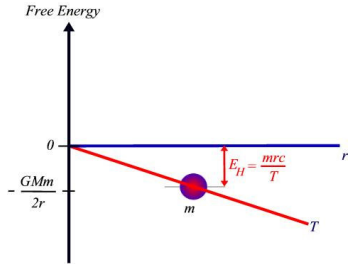

**Figure 5. A situation where a mass in a gravitationally stable orbit may spontaneously become free**

In this free energy diagram, the condition of interest is:

$$\frac{Gm}{r^2} = acceleration = \frac{2c}{T} = 2a_o \qquad (5)$$

We have neglected the relativistic boost on $E_H$ in the transition between reference frames. The notation $a_o$ is used to represent $c/T$. In moving from bound to free state there is no energy penalty and we might expect a spontaneous transition to occur. There is certainly some evidence of curious behaviour around these acceleration levels. This model suggests the reason for this is not a breakdown in gravity, but because there exists more than one spacetime.

Up to now we have considered only the special case where the free mass has no proper motion. What if the mass gains more than the minimum energy to escape the binding force(s)? The excess will appear as the kinetic energy of proper motion. With proper motion in addition to the expansion velocity, the Hubble velocity will no longer remain constant in the local frame. If we consider the case of a free mass with proper velocity $v_P$ (where the positive direction is naturally away from the observer), then the movement relative to the cosmological frame results in a change in apparent energy. In Fig. 4, this is equivalent to a component of motion parallel to the red (sloping) line.

Proper motion cannot therefore be incorporated into the local frame in a consistent way. An anomalous force appears that can be identified with an interaction between the two spacetimes.

Differentiating the energy difference between the spacetimes with respect to $r$, gives the apparent force associated with the effect from the viewpoint of the bound observer. Again we neglect the relativistic correction. From (4) and remembering that the energy difference was defined as positive, but is negative in the observer frame,

$$-\frac{dE_H}{dr} = -\frac{m_o c}{T}\left(1 - \frac{r}{T\frac{dr}{dT}}\right) \qquad (6)$$

If we let $dr/dT$ equal $v_P + v_H$ and using (2), $v_H = r/T$, the anomalous acceleration can be defined as $\psi$:

$$\psi = -a_o\left(\frac{v_P}{v_P + v_H}\right) \qquad (7)$$

This is an apparent retardation to the motion that works to eventually convert the proper motion completely into a recession velocity.

All free mass exhibiting proper motion will be subject to this virtual force. It is possible to interpret this as some sort of frictional effect between the two spacetimes but such analogies are probably inadequate and unnecessary. In look at energy differences between spacetimes, it is meaningless to ask where the energy goes, but note that if the direction of proper motion is reversed, and is of greater magnitude than the Hubble velocity, energy is recovered[xv].

7. TESTS

If the expansion does not act locally *and* space is substantial then there must be two coexisting spacetimes about us. There is no alternative. We have chosen the simplest relationship between frames that gives a relativistic and energy consistency, a basic framework that is readily extended to general relativity[xvi]. The theory makes measurable predictions. A free mass with proper motion will be subject to a deceleration with respect to the observer of the order of $a_o$ when $v_P \gg v_H$.

Earth based experiments could test this. A mass or particle is free if it exceeds the solar escape velocity – it does not actually have to escape the Earth. There are very few examples of masses that have escaped the gravitational constraint of the solar system. There are several spacecraft, particularly the Pioneer 10 and 11 probes. These have been subject an anomalous acceleration of the same direction and magnitude as that predicted by (7). Taking the age of the Universe (*T*) as 13 billion years, the predicted acceleration is 7.3 x $10^{-10}$ ms$^{-2}$ [37], rather lower than the value of 8.8 x $10^{-10}$ ms$^{-2}$ derived from the data. However, this may be pointing to the failure of the simple model present here to incorporate GR. The real test would be to analyse the Pioneer data at the transition point from bound to unbound and look for a characteristic signature[xvii].

The second prediction is that when the acceleration of orbitally bound bodies drops to $2a_o$ then the mass is no longer bound and becomes subject to the Hubble expansion. The anomalous behaviour of stars in spiral galaxies appears to be associated with accelerations of the order of $a_o$ [38].

## 8. DISCUSSION

It is tempting to further hypothesise that the gravitational constant changes with time, but only in the global frame. Thus a change in the gravitational constant cannot be detected by solar system experiments[xviii]. One attractive possibility is a gravitational constant increasing in proportion to *T*. This offers a natural explanation for the flat rotation curves that appear typical of most spiral galaxies. Consider a star in the outer regions of a galaxy that was bound in a stable orbit but suddenly makes a spacetime transition (as shown in Fig. 5) becoming subject to the cosmological expansion. The star will forevermore maintain a constant radial velocity *v* if

$$\frac{d}{dr}\left(\frac{G}{r}\right) = 0 \quad (8)$$

because $v^2$ is proportional to *GM/r* ( assuming the star mass loss is minimal). This is equivalent to the condition

$$\frac{dG}{dt} = \frac{G}{T} \quad (9)$$

i.e. G ∝ T. A gravitational constant varying in proportion to *T* will automatically result in flat galactic rotation curves.

We can picture a spiral galaxy developing from material escaping the grip of the central bulge and moving outwards under the cosmological expansion. The Tully-Fisher relation emerges naturally as well [38]. Equating (5) and $v^2=GM/r$, we get the radial velocity when the material escapes the core:

$$v^4 = \frac{G}{T}Mc \quad (10)$$

Although material can escape at any time in the history of the galaxy, *G/T* is always constant, thus we obtain the Tully-Fisher relation (assuming as is conventional that mass is proportional to luminosity). This is identical to the MOND expression (which is known to explain the rotation curves of spiral galaxies), except their $a_o$ is a factor of 6 smaller than ours [38]. However, (10) refers to the core luminosity only, not the galaxy luminosity as in MOND.

Fig. 6 shows an idealised rotation curve for a barred spiral galaxy. The central nuclear bulge executes rigid body rotation (*v* ∝ *r*); the surrounding less-dense material shows Keplerian motion (*v* ∝ $r^{-1/2}$) out to the constrained limit (gravitational acceleration ≈ $2a_o$) ; then a mostly steady velocity equal to the radial velocity at the time of escape (generally greater further out because of the increased central mass at that stage). Of course, the individual segments are not really that well-defined – the diagram also shows a smoothed version incorporating the variation due to the local mass distribution. This is a perfectly standard rotation curve, very similar to that of our own Milky Way galaxy or the galaxy NGC 2590 [39]. This is of course phenomenological only but the flat-ish rotation curves that are prevalent are explained very simply by a changing gravitational constant without invoking Dark Matter or MOND, both of which are known to be problematic[xix].

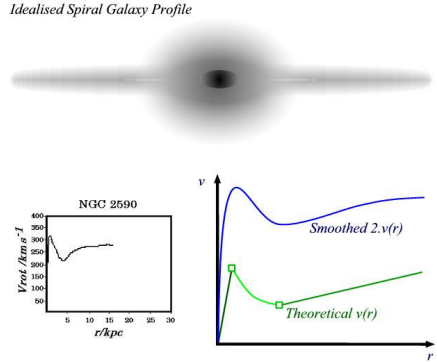

**Figure 6. The theoretical rotation curve of a typical galaxy. The piecewise construction is smoothed and is shown to be comparable with galaxy NGC 2590.**

Another point about a spacetime with the gravitational constant increasing linearly with time is that the expansion does not do nett work against the gravitational field – the work done is exactly balanced by the increase in the gravitational potential. The problem with this is that the global frame is not then described by the conventional Friedmann equations - there is no braking effect and the expansion proceeds independently of the mass distribution through the Universe. A significant modification to the Friedmann equations would be required.

An unusual consequence of a linear increase in the gravitational constant is the gravitational force becomes scale independent in the global frame. The force between two receding objects is thus constant with time. In effect, as far as gravitation is concerned, the global spacetime frame appears static.

## 9. SUMMARY AND CONCLUSION

We have demonstrated that the failure of the cosmic expansion to act locally, and the substantial nature of spacetime imply the proximity of two coexisting spacetimes in the solar neighborhood. The assumption was made that all matter is attached to one or other spacetime. Of interest was how mass associated with one spacetime would appear when viewed from the other, given that the spacetimes may not share the same energy reference. By demanding that all observation in a frame be consistent with the salient characteristics of that frame, a unique equation was derived to describe the energy reference difference between spacetimes. This equation predicts that unbound masses with proper motion in addition to the recession velocity will experience a constant deceleration when observed from the local frame, and the deceleration will be in the direction of the centre of mass. It is also predicted that stable gravitationally bound systems cannot extend beyond the point where the gravitational acceleration drops below the level ~ *c/T*. These are testable predictions. One can imagine astronauts being accelerated to escape velocity and moving from our local spacetime to the global spacetime.

A change in the gravitational constant over time has been proposed many times, but has always been rejected because a change was never detected by experiments conducted within the solar system. With two spacetimes instead of one, it is possible that the 'constant' changes in one spacetime but not the other. Though highly speculative, the consequences of an increasing gravitational constant in the expanding frame are considered, and found to provide some qualitative insight into the properties and development of spiral galaxies. The proposal is completely consistent with general relativity.

The idea introduced in this paper is simple and the calculations elementary. Nevertheless the proposal may have some merit, but is really only applicable to observations and tests at the extremities of galaxies. It will be based on information about this boundary region that the proposal of coexisting spacetimes will ultimately stand or fall.

NOTES

[i] The terminology on this subject is loose – the term 'parallel universe' is used in many ways and does not necessarily refer to a background universe 'close' to our own.

[ii] If analogy is needed, this is rather like a dividing cell where the DNA is copied into the daughter cells, but always with the chance of a mutation.

[iii] We can legitimately asked how the process is managed and controlled. How are errors prevented? Is there somewhere in here a tantalising clue about the identical nature of elementary particles through this perfect duplication process? Supporters of the MWI do not generally explore the mechanism by which the splitting takes place, but this is certainly a legitimate line of enquiry, otherwise the questions about how the world operates are simply pusher to a deeper less accessible level.

[iv] Of some interest is how Mach's principle deals with an expanding Universe. If the inertial reference consists of stars in distant galaxies, they are not stable references as Mach imagined.

[v] Using whatever the notion of simultaneity is in the MWI.

[vi] The theories are unclear about cases where there may be a deficit of future absorbers; are photons still emitted?

[vii] The laws of physics then just reveal or predict the future.

[viii] Without the big bang model, the picture is straightforward (although we have no model to explain why the expansion should be occurring at all). The distance between galaxies is expanding linearly with time. The photons move along straight lines not curved paths - parallel lines do not cross. With inflation, new puzzles arise. Statements such as 'Opposites edges of the observable Universe are 28 billion light-years separate, yet they look similar (the uniformity of the cosmic background). How is this possible when there is no causal connection?' are often presented [17]. Without inflation, this is not an issue. We see each side of a region very much smaller (when the Universe really was smaller) and in thermal equilibrium at the time of emission.

[ix] In many ways this topic is clouded by too great a focus on the way the parameters we appear to be landed with in this Universe give rise to the necessary complexity. Other parameter set choices, with a full analysis may be found to give rise to a Universe of equal complexity and a different form of intelligence. It might be that we can never understand the Universe as a whole, that we are not capable of understanding it, thus the anthropic approach can never be shown to be wrong. After all, the human intellectual goal is to reduce the Universe to null [24]. But it is not null. We find it impossible to understand why anything should exist at all. But things do exist. Conservation principles are deeply embedded in the way we think, and this may be the basis for mathematics, but this may not be sufficient to understand the entire Universe.

[x] The sheer lack of imagination in using oscillating wave analogies for sub-particle processes is discouraging – one would hope for wave effects to emerge from a fundamental theory rather than be introduced.

[xi] String theory takes the idea that there is three space dimensions *as is*. It really makes no effort to explain why three? This is a valid line of enquiry because the number of dimensions does not change with time in the current epoch. We may project backwards in time to the stage where there were no particles and the number of dimensions is meaningless. By what process could three dimensions emerge as particles were being introduced? If one is to adopt a relationist view of spacetime, this problem must to be solved.

[xii] It is meaningless to ask how the two substantive spacetimes can exist together 'in the same place'.

[xiii] Why the centre of mass? No other interpretation can give a consistent ordering of events when views from various positions in the observer frame.

[xiv] Equation 4 states the apparent mass of a particle tends to zero as we look further and further back, becoming zero when the recession velocity is $c$, at $T = 0$ in our simple model. In other words, this energy is associated with rest mass energy. These equations were derived before in the context of a real energy in a single spacetime [35], however in a single spacetime things do not quite hang together. Note that there is an overlap reminiscent of hysteresis that ensures there is not a 'no man's land' – a mass will always be bound to one or other spacetime based on energy. There is no third spacetime possible.

[xv] The model does not really explain the loss of energy associated with the redshift, and it is reasonable to ask where the missing energy goes. However, we are working between two separate spacetimes when the notion of energy conservation is very much blurred (although no doubt it holds absolutely). There is the related idea of why the energy is not recovered when mass is converted to energy. The photon received is still redshifted. However, the received photon is still technically unbound.

[xvi] But note that GR has only been proven to be effective in the local frame.

[xvii] As will be seen in the next section, the solar system is probably expanding with respect to the galactic core. The centre of mass of our local spacetime is therefore located close to the sun. We would expect the Pioneer acceleration to be directed towards the sun, not the earth or the galactic centre. There are many local spacetimes. We are not part of the galactic local spacetime.

[xviii] Arbab [36] as considered this possibility that the gravitational constant increased over the geological history of the Earth, but the concept, no matter how attractive seems flawed. Tests in the solar system have found no changes in the planetary distances consistent with a change in the gravitational constant in the order of that predicted. It should be noted that there is no possibility that the cosmological expansion balances a local change in the gravitational constant. The numbers simply are not comparable. This is all very speculative of course, but remember that we have two spacetimes. It is conceivable that G changes in one but remains constant in the other. After all there is a definite connection between expansion and the change in the gravitation constant as described in Section 5.

[xix] The dynamics are more complex, but we can certainly propose that as material is ejected from the core via jet action, material at the peripheries that was previously bound escapes and is subject to the Hubble flow. The velocity is of the order of 0.5 km/s for the sun but remains constant as previously suggested. For the sun to be in the current position, the material would have left the centre when the Universe was a third of its current age. Of course, there is further ejecta at the ends of the bar moving at a greater speed and this more concentrated material may be the origin of the spiral arms (there is evidence that arms originate from the end of bars).